# TRANSVERSE MODES FOR FLAT INTER-BUNCH WAKES*

A. Burov, FNAL, Batavia, IL 60510, USA

*Abstract*

If inter-bunch wake fields are flat, i.e. their variations over a bunch length can be neglected, all coherent modes have the same coupled-bunch structure, provided the bunches can be treated as identical by their inner qualities (*train theorem*). If a flat feedback is strong enough, the transverse modes are single-bunch, provided the inter-bunch wakes are also flat (*damper theorem*).

## TRAIN THEOREM

When bunches are equidistant along a circular orbit, a coupled-bunch (inter-bunch) structure of beam collective modes is known to be independent of their head-tail (intra-bunch) structure. Indeed, in this case the coupled-bunch mode behaviour is universally described by an exponential factor $\exp(-i\mu_n k)$, where $k$ is a bunch number, $\mu_n = 2\pi n / M$ is an inter-bunch phase advance, with integer coupled-bunch mode number $0 \leq n \leq M-1$ and $M$ as the total number of bunches. The inter-bunch mode classifier, phase advance $\mu_n$, is identical for all the head-tail mode parameters, universally describing all the head-tail modes. For the equidistant bunches, this inter-bunch mode universality is valid for any wake function; it is just a direct consequence of the beam translation invariance. Note that this inter-bunch mode structure is preserved for colliding beams as well, provided that both beams are translation-invariant, i.e. their bunches are equidistant. Normally, multi-bunch beams in circular machines are not translation-invariant. As a consequence of injection limitations, they a have more complicated bunching structure, consisting of some clusters of bunches, normally called trains or batches. Also note that if the trains are equidistant, they can be treated as equidistant super-bunches, and the problem is reduced from the full number of bunches to the train number of bunches.

If the beam is not translation-invariant, the inter-bunch mode structure is generally sensitive to the intra-bunch one; the former cannot be disentangled from the latter. However, there is one special and practically important case, when the inter-bunch mode structure is independent of the head-tail one, without any requirement for a batch/train structure of the beam.

Let's first consider beam bunches as $M$ macroparticles arbitrarily distributed along the orbit. Knowing the inter-bunch wake function $W_{kl}$, $0 \leq k,l \leq M-1$, all $M$ coupled-bunch modes can be found by solving a conventional eigensystem problem

$$\Omega X_k = W_{kl} X_l. \qquad (1)$$

Note that here inter-bunch wakes and fields left behind by the same bunch at previous revolutions can be taken into account, providing identical diagonal elements of the wake matrix **W**. Thus, for this simple model, all $M$ coupled-bunch tune shifts $\Omega_n$ and eigenvectors $\mathbf{X}_n$ can be easily found.

Now, let's take into account a finite bunch length and all possible intra-bunch modes, assuming that the inter-bunch wake is flat, meaning that its variation over the bunch length can be neglected. This situation is typical: when the high-Q high-frequency parasitic cavity modes, as well as electron clouds, are sufficiently suppressed, this is the case, provided the bunch length is small compared with a bunch separation. For the flat wakes, only positions of the bunch centroids matter; all other details of the intra-bunch motion do not count. Therefore, for the inter-bunch motion, the bunches are equivalent to macroparticles, and so all the head-tail modes have the same inter-bunch structure given by the eigenvectors of Eq. (1), Q.E.D. This statement, referred to as the *train theorem*, assumes flat inter-bunch wake function, and is valid for an arbitrary batch/train sequencing of the beam. Whenever it is applicable, the train theorem suggests an extremely powerful tool, reducing a multi-bunch beam dynamics problem to a single-bunch one. Indeed, this reduction is straightforward. First of all, a trivial eigensystem problem of Eq. (1) has to be solved; its solution gives the coupled-bunch mode structure. As soon as it is done, the state of every bunch in the beam can be presented as similar to the state of a reference bunch, with the similarity ratio given by a ratio of the two corresponding components of the eigenvector **X**. After that all the wake fields left behind by the preceding bunches are expressed as functions of a state of the reference bunch; thus, the multi-bunch problem is reduced to a single-bunch one. This problem reduction is extremely powerful for a beam with a large number of bunches; it is equally useful both for tracking codes and Vlasov solvers. The train theorem shows all the coupled-bunch modes as independent, allowing parallel analysis of ~10-20 of their representatives.

Note that this inter-bunch mode universality does not mean that the intra-bunch motion is not sensitive to the inter-bunch mode number; generally, the intra-bunch modes depend on the inter-bunch mode number.

Is the train theorem applicable to colliding beams? At first glance, the answer is positive, provided the bunch length is much smaller than the beta-function in the collision point. Indeed, in this case the beam-beam coherent interaction is reduced to a cross-talk between their centroids, so its "wake" is flat, and the theorem can

---

be applied. However, in many cases there is still a significant obstacle to its application, associated with long-range parasitic collisions happened before and after the main interaction point. Although the long-range wakes are normally very flat (i.e. the related beta-functions significantly exceed the bunch length), the other condition of the train theorem is broken by the long-range collisions: bunches cannot be treated as inner-identical any more. The long-range collisions introduce a difference between incoherent tune shifts of different bunches: train-edge ("pacman") bunches have less long-range collisions than the regular ones. Therefore, Landau damping for the pacman bunches is not the same as for the regular ones; the bunches are significantly different, and so the train theorem is not applicable. Thus, with long-range collisions of multi-bunch beams, beam stability analysis appears to be extremely complicated: the bunches are significantly different, so a complexity of the single-bunch problem is multiplied by the complexity of the coupled-bunch one.

However, there is an elegant way out of this obscurity.

## DAMPER THEOREM

Normally circular accelerators with intense multi-bunch beams are equipped with transverse dampers whose bandwidth is comparable with a frequency of the bunches. A damper of such a kind detects bunch centroids and reacts on them, being unable to resolve other details of the intra-bunch motion; in other words, its wake is flat. In case its damping rate, the gain, exceeds impedance-related coherent tune shifts, the damper becomes one of the main factors determining both the coherent tunes and the intra-bunch structures of the head-tail modes. This asymptotic case is of a special interest, since it shows ultimate capability of the damper, presumably reached when its gain is high enough.

What happens with the bunch modes when the gain is getting higher and higher? Every head-tail mode has its amplitude of the bunch centroid oscillations. Since the damper does nothing but seeing and reacting on that amplitude, it provides the given mode with a damping rate proportional to the centroid amplitude of this mode. Thus, with sufficiently high gain, every mode with non-zero centroid amplitude is going to be seen and damped by the damper; all the unstable modes are those with zero centroid amplitudes and therefore invisible to the damper. However, these potentially unstable zero-centroid modes are invisible not only to the damper. They are invisible to other bunches as well. Indeed, for the flat wakes, the bunches interact with each other by means of their centroids only. If the centroid motion is blocked, they cannot react to each other, and this is true both for the bunches of the same and of the opposite beam. With a sufficiently strong damper, both coupled-bunch and coupled-beam oscillations are turned off, and so the bunch coherent oscillations are invisible anywhere outside this very bunch [1]. This is the damper theorem, assuming flat wakes only. Contrary to the train theorem, the damper theorem does not set any requirement on the bunch similarity; thus, the damper theorem is applicable for any collision scheme, with or without long-range collisions, provided that the bunch length is short compared with the beta-function. Therefore, the damper theorem reduces multi-bunch beam dynamics problem to the single-bunch one for arbitrary storage rings or colliders, requiring the wake flatness only. Note that weak-strong (incoherent) beam-beam effects are not excluded by this theorem: for every individually oscillating bunch, nonlinear focusing of the ongoing beam should be taken into account for a proper computation of Landau damping.

The damper theorem is asymptotic one, it tells something for a case when one of system parameters, the damper gain, is "sufficiently high". How high should the gain be to make the damper theorem practically applicable? Qualitatively, the answer appears to be rather obvious. A tune shift parameter for the coupled-bunch oscillation is given by a maximal eigenvalue of Eq. (1). Thus, to suppress the coupled-bunch oscillations, the gain has to exceed sufficiently by the absolute value all the coupled-bunch tune shifts. Similarly, a parameter for the coupled-beam oscillations is the coherent beam-beam tune shift. Thus, to exclude these oscillations, the gain has to exceed, by the absolute value, the strong-strong beam-beam tune shift. However, unlike the coupled-bunch tune shifts, the coupled-beam ones are purely real. That is why, a suppression of the coupled-beam oscillations does not require the gain to be much higher than the beam-beam tune shift; it is already sufficient for the former to be comparable to the latter [1-3].

## SUMMARY

For many practical cases, a problem of multi-bunch collective stability in a circular machine is reduced to a single-bunch problem: either by means of the train theorem, or by means of the damper theorem.

## ACKNOWLEDGMENT

I am extremely thankful to Elias Metral, my CERN host during my FNAL-LARP long-term visit to CERN – not only for his permanently warm hospitality, but also for innumerable discussions, which helped me to see the main areas where my efforts could be mostly effective for the LHC complex. I am also grateful to Stephane Fartoukh, Nicolas Mounet and Simon White for multiple useful discussions associated with a content of this paper.